An Air-Stable and Atomically Thin Graphene/Gallium Superconducting Heterostructure

Brian Bersch[1,2], Natalie Briggs[1,2], Yuanxi Wang[3], Jue Jiang[3], Ke Wang[4], Chengye Dong[1,2], Shruti Subramanian[1,2], Mingming Fu[5], Qiang Zou[5], Ya-Wen Chuang[3], Zheng Gai[5], An-Ping Li[5], Jun Zhu[3], Cui-Zu Chang[3], Vincent H. Crespi[3,6], Joshua A. Robinson[1,2]*

[1]Department of Materials Science and Engineering, The Pennsylvania State University, University Park, PA 16802, United States

[2]Center for 2-Dimensional and Layered Materials, The Pennsylvania State University, University Park, PA 16802, United States of America

[3]Department of Physics, The Pennsylvania State University, University Park, PA 16802, United States

[4]Materials Research Institute, The Pennsylvania State University, University Park, Pennsylvania 16802, United States

[5]Center for Nanophase Materials Sciences, Oak Ridge National Laboratory, Oak Ridge, TN 37831-6487, United States

[6]Department of Chemistry, The Pennsylvania State University, University Park, PA 16802, United States

*Corresponding author

Email: jrobinson@psu.edu

**Two-dimensional layered**[1–5] **and atomically thin elemental**[6–9] **superconductors may be key ingredients in next-generation quantum technologies**[10]**, if they can be stabilized and integrated into heterostructured devices under ambient conditions. However, atomically thin elemental superconductors are largely unexplored outside ultra-high vacuum due to rapid oxidation, and even 2D layered superconductors require complex encapsulation strategies to maintain material quality**[11]**. Here we demonstrate environmentally stable, single-crystal, few-atom-thick superconducting gallium, 2D-Ga, produced by confinement heteroepitaxy (CHet) at the interface of epitaxial graphene (EG) and silicon carbide (SiC). 2D-Ga becomes superconducting at 4 K; this elevation over bulk α-Ga ($T_c$~1 K)**[12] **is attributed to an increased density of states at the Fermi level as the incipient Ga-Ga dimerization seen in α-Ga is suppressed by epitaxy to SiC. This work demonstrates that unique 2D forms of 3D materials can be stabilized at the EG/SiC interface, which represents a scalable route towards air-stable crystalline 2D superconductors as a potential foundation for next-generation quantum technologies.**

Tremendous advances in fundamental science have followed from the burgeoning practice of exfoliation, stacking, and encapsulation of diverse atomically thin 2D layers into functional heterostructures at the micron scale[11]. The next step towards technological impact is to transition the functional diversity of these highly sophisticated "pick and place" devices to a *wafer-scale* platform. Additionally, the sensitivity of 2D systems to environmental influences – largely addressed at the micron scale by hBN encapsulation of individual devices[13] – poses additional challenges to technological deployment, particularly for the more reactive metallic or small-gap semiconducting monolayers that host much of the compelling new physics arising from strong spin-orbit coupling or topological superconducting order[4,14]. Since heterolayer growth in multilayer devices is a particularly invasive "environmental influence" (i.e. through interfacial reactions), this sensitivity presents a profound challenge to capturing the promise of 2D quantum materials in scalable platforms. A general platform for producing environmentally stable wafer-scale 2D-superconductors with relatively high transition temperatures would represent a significant step towards this goal. Here we describe such a process; naturally encapsulated 2D-Ga formed at the EG/SiC interface exhibits a superconducting phase with critical transition temperature ($T_c$) of 4 K.

Formation of air-stable 2D-Ga utilizes monolayer epitaxial graphene grown by silicon sublimation from 6H-SiC(0001) substrates with two primary SiC step-edge morphologies: "large-step" and "small-step" (**Methods, Figures S1, S4**). Following graphene formation, EG is exposed to a low-power oxygen plasma to introduce a uniform distribution of defects that act as "landing sites" and pathways for Ga intercalation through the graphene to the EG/SiC interface[15,16]. Uniform 2D-Ga layers are subsequently formed by thermal evaporation of metallic Ga at 800°C (**Methods**, **Figure S2**). Importantly, following intercalation, the graphene Raman D band (whose intensity is very sensitive to the concentration of structural defects in the basal plane) shrinks to nearly the same intensity as in as-grown EG (**Figure S5**), indicating that the graphene overlayer heals. X-ray photoelectron spectroscopy demonstrates healed EG forms a hermetic seal that enables facile *ex situ* characterization of 2D-Ga and prevents oxidation even 9 months after synthesis (**Figure S3**), which otherwise would rapidly degrade in air. We attribute this "healing" to a combination of high-temperature annealing and Ga-catalyzed graphene regrowth[17]. This elemental intercalation and stabilization process, dubbed Confinement Heteroepitaxy (CHet), creates 2–3 atom thick 2D-Ga (**Figure 1a, S4**) that is epitaxial to the underlying SiC[15]. Scanning electron microscopy, Auger electron spectroscopy (**Figures 1b-c, S7**), and Raman mapping (**Figure S6**) confirm that oxygen-plasma treatment prior to intercalation is required to achieve laterally uniform 2D-Ga. Normalized resistance vs temperature (R(T)) measurements (**Figure 1d**) confirm that only O$_2$-plasma-treated EG, when Ga intercalated, yields a complete superconducting transition. See **Methods** for details on *ex situ* electrical measurements. Ga-intercalated *as-grown* EG (i.e. without plasma treatment) exhibits a >10× higher normal-state resistance at 5 K and a slight drop in resistance at ~4 K with large residual resistance ~600 Ω at 2 K (**Figure 1d** inset), suggesting that localized 2D-Ga domains may superconduct but do not yield macroscopically superconducting films. In the case of intercalated as-grown EG, a continuous superconducting path is not formed between electrodes due to non-uniform and inefficient Ga intercalation through pristine graphene. Thus, all further discussion focuses on 2D-Ga films synthesized using plasma-treated graphene.

Scanning tunneling microscopy (STM) and scanning tunneling spectroscopy (STS) demonstrate that the superconducting energy gap is not strongly affected by nanoscale topography, including the presence of nanometer-size steps in graphene/SiC(0001) and defect sites within terraces (**Figure 1e-g, S8**) that may be byproducts of the O$_2$-plasma treatment and subsequent annealing/intercalation. Local differential tunneling conductance (dI/dV) spectra (**Figure 1g-i**) reveal a consistent superconducting energy gap at multiple locations on the sample (three of which are shown in **Figure 1e**), independent of the local topography of the graphene overlayer. A superconducting energy gap with well-defined coherence peaks at ±0.6 meV is measured at 2.2 K (**Figure 1g**) and gradually diminishes with increasing temperature (**Figure 1h**). The disappearance of the energy gap at ~3 K in STS (i.e. somewhat lower than seen in transport) may be due to a reduced magnitude of a proximity-induced superconducting order parameter in the 2–3L graphene overlayer[18]. The dI/dV spectra exhibit an increase in the zero-bias conductance and the disappearance of coherence peaks with increasing perpendicular B up to 0.1 T (**Figure 1i**). Although spatially rare, partially intercalated Gr/Ga terraces with islanding of ~1L-Ga do not yield a clear signature of superconductivity down to 2.4 K in STS (**Figure S9**), which we attribute to a combination of reduced Ga thickness and greatly reduced average domain size compared to the coherence length in 2D-Ga.

Substrate surface morphology, in particular crystallographic steps in the SiC(0001) surface that form during graphene synthesis (i.e. step bunching), strongly affects long-range continuity of the intercalant in 2D-Ga and thus also affects carrier transport, as revealed by R(T) measurements for contacts oriented parallel and perpendicular to these steps (**Figure 1j-k**). In the perpendicular configuration, field lines and current paths must traverse the steps, while this is minimized in the parallel configuration. In the case of "small-step" EG/SiC (steps ~1 nm tall), R(T) both parallel and perpendicular orientations display a superconducting transition with similar $T_c$(onset)~4 K and $T_c$(zero)~3.2 K, with only a modest 20% increase in the normal-state resistance (and a slight secondary transition at ~3.5 K) for the perpendicular case. On the other hand, "large-step" EG/SiC (steps 5–20 nm tall) shows distinct transitions for the two configurations: while both have a $T_c$(onset)~3.8 K similar to that of the "small-step" EG/SiC, only parallel transport displays a fully developed superconducting transition with $T_c$(zero)~2.5 K, suggesting that transport perpendicular to large steps cannot access a continuous current path through superconducting material and instead encounters a finite series resistance at the steps (**Figure S4**). Thus, epitaxial graphene synthesis must be optimized to limit step bunching across the SiC (0001) surface to ensure uniform superconducting films with isotropic transport.

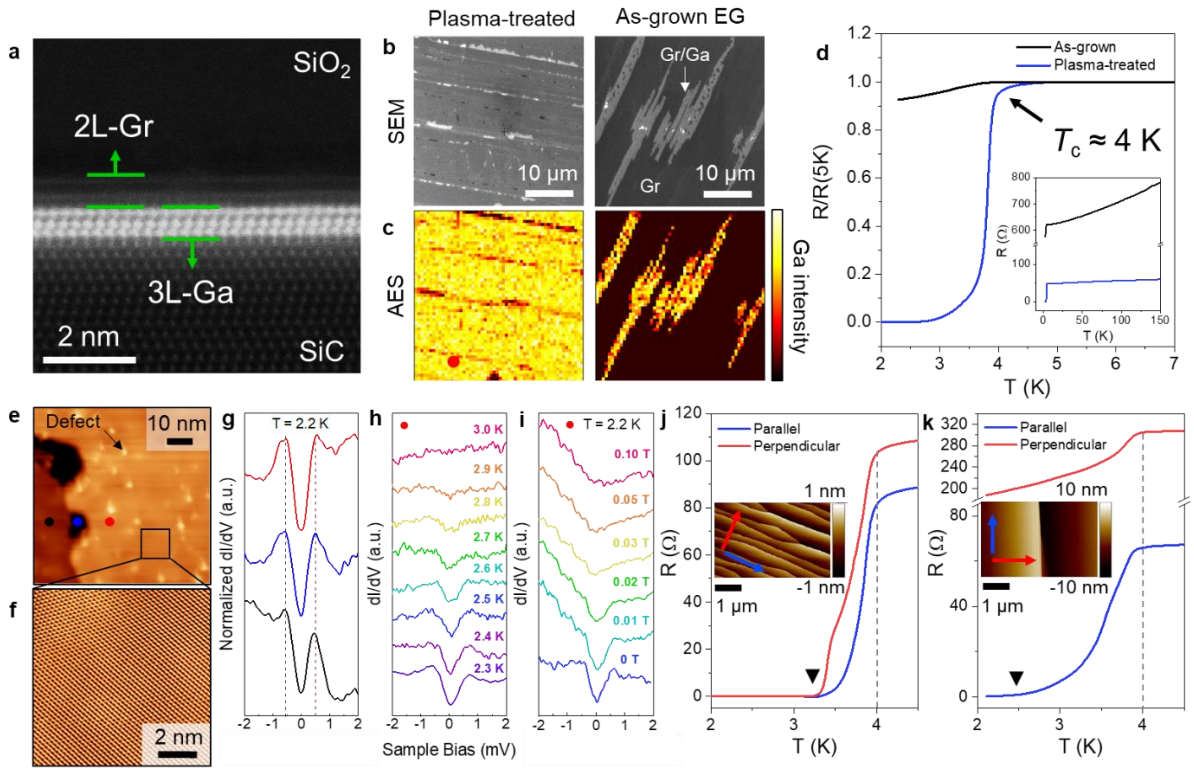

**Figure 1| Characterization of graphene/2D-Ga heterostructures. a**, High-resolution scanning transmission electron microscopy (HR-STEM) of EG/2D-Ga. **b-c**, Scanning electron microscopy and auger electron spectroscopy maps of 2D-Ga intercalated using plasma-treated and as-grown EG. **d**, Resistance vs temperature (R(T)) plot (normalized to resistance at 5 K) for EG/Ga heterostructures formed from plasma-treated and as-grown graphene. Figure 1d inset: Non-normalized R(T) curves for the same samples from 2 to 150 K. **e-f**, Large-area and atomic resolution scanning tunneling microscopy (STM) images of the EG surface topography, respectively. **g**, Differential conductance (dI/dV) spectra (normalized to 3.0 K spectra) taken at 2.2 K for the 3 different regions shown in the STM image in **e**. **h-i**, Temperature-dependent (zero-field) and perpendicular magnetic-field-dependent (2.2 K) dI/dV spectra, respectively, of the right-terrace in **e**. **j-k**, R(T) plots comparing perpendicular and parallel current directions performed on small-step and large-step samples (both plasma-treated). Black arrows indicate approximate $T_c$(zero) values, and dashed lines are meant to aid the eye. The curves in **d** were measured on a large-step sample in parallel configuration.

R(T) for small-step 2D-Ga from 300 K to 2 K (**Figure 2a**) shows largely metallic behavior down to 4 K, below which a superconducting transition to a zero-resistance state is observed. A log plot from 5 K to 2 K shows a sharp four-order-of-magnitude drop in resistance between normal and superconducting states (**Figure 2a** inset). 2D-Ga exhibits a $T_c$(onset)=3.95 K, $T_c$(0.5$R_N$)=3.83 K, $T_c$(zero)=3.2 K, and thus a transition width of $\Delta T_c$≈0.75 K, where the various $T_c$ criteria are defined in **Methods**. As expected, $T_c$ decreases and the transition broadens with increasing perpendicular magnetic field ($B_\perp$) (**Figure 2b**). R(B) measurements (**Figure 2c**) indicate an out-of-plane critical field of $B_{c2}$=130 mT at 2 K and a corresponding coherence length of ξ~50 nm (**Methods**). A linear extrapolation of $B_{c2}$(T) data extracted from R(B) curves in **Figure 2c** indicates a zero-Kelvin critical field $B_{c0}$≈260 mT and corresponding coherence length ξ$_0$~36 nm, higher than that of α-Ga ($B_{c0}$≈6 mT)[12] and β-Ga ($B_{c0}$≈54 mT)[19] (**Figure S12**). The Berezinskii–Kosterlitz–Thouless (BKT) transition temperature is extracted for both parallel and perpendicular current directions (**Figures 2d-e**), where 2D superconductivity is characterized by a transition from V ∝ I in the normal state to V ∝ $I^\alpha$ in the superconducting state, and the temperature at which α=3 is defined as $T_{BKT}$[20] (**Figure 2f, Methods**). Here, $T_{BKT}$=3.1 K (2.9 K) for the parallel (perpendicular) configuration in small-step 2D-Ga. Although the actual $T_{BKT}$ is likely higher if the power law exponents are curve-fitted closer to the critical current $I_c$[6,20], the similar $T_{BKT}$ observed for both current directions indicates nearly isotropic transport in 2D-Ga/SiC. A plot of $d\ln(R)/dT)^{-2/3}$ vs T indicates a $T_{BKT}$=3.88 K (**Figure S12**), which reinforces the observation that the $T_{BKT}$ values extracted from the I-V curves in **Figures 2d-e** represent lower-bound estimates and are limited by the measurement setup (**Methods**).

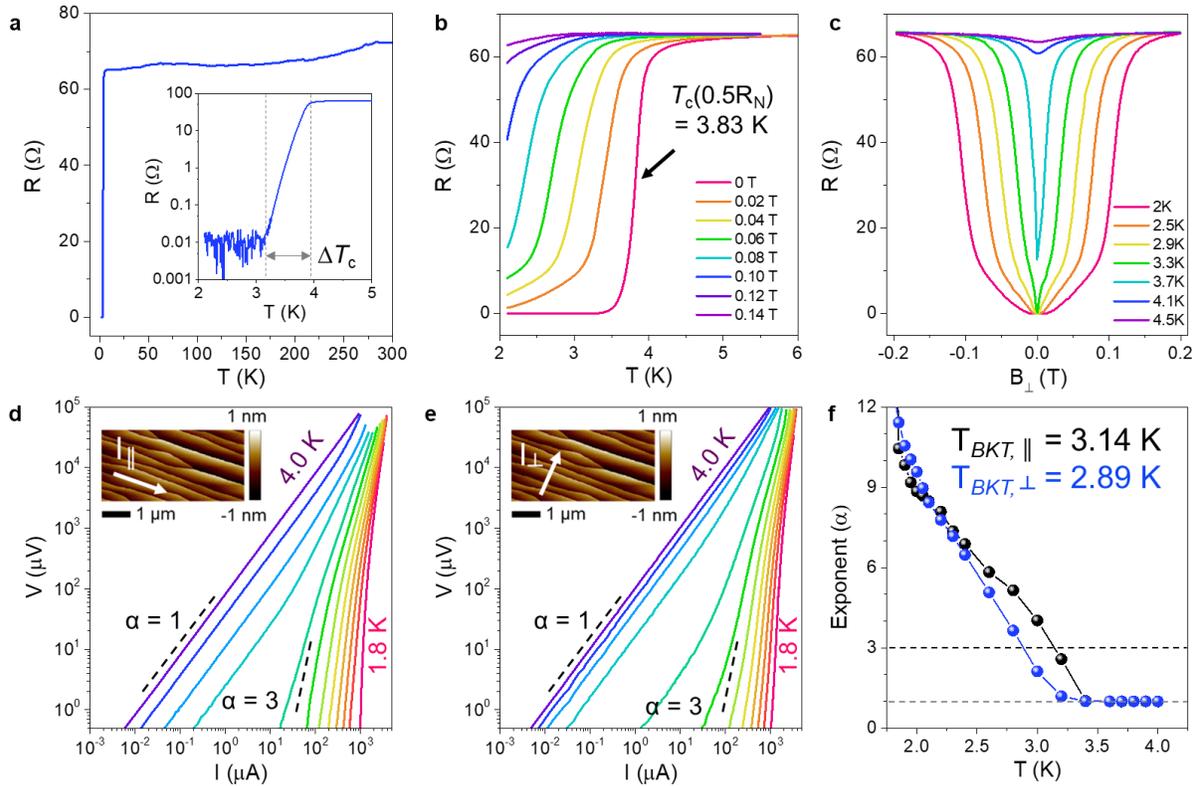

**Figure 2| Transport results of graphene/2D-Ga heterostructures. a**, Zero-field R(T) curve for an optimized 2D-Ga (plasma-treated, small-step, parallel configuration) film from 300 to 2 K. Inset of **a**: Log-scale plot from 5 to 2 K of the same curve. **b**, R(T) curves showing a degradation in $T_c$ and increase in residual resistance at 2 K with increasing out-of-plane magnetic field. **c**, Resistance vs out-of-plane magnetic field (R(B)) curves showing a similar degradation in the superconducting state with increasing temperature. **d-e**, Current-voltage (I-V) curves measured in parallel and

perpendicular current directions, respectively, on the same small-step sample. **f**, Exponent (α) vs temperature plots for both measurement directions displaying the lower-end estimates of the BKT transition temperatures.

The onset temperatures for superconductivity in hexagonal CHet-based 2D-Ga/SiC (4 K) and 2L-Ga/GaN (5.4 K, produced in UHV by direct deposition, not intercalation)[6,9] are higher than bulk α-Ga (orthorhombic, $T_c$=1.08 K)[12], similar to metastable β-Ga (monoclinic, $T_c$=5.9–6.5 K), and below metastable amorphous Ga ($T_c$=8.4 K)[19,21,22]. Naively, hexagonal 2D-Ga appears to more closely resemble β-Ga structurally than α-Ga (**Figure 3a**). To interrogate this intuition and verify that superconductivity originates from Ga (and not graphene), we calculate in first-principles density functional theory the electronic densities of states (DOS) near the Fermi level for all three systems (**Figure 3b,c**) and also the Eliashberg spectral function $\alpha^2F(\omega)$ for three-layer Ga epitaxial to SiC (without the graphene cap, see **Methods**). The α-Ga lattice contains Ga dimers; this incipient covalency produces a dip in the DOS. In contrast, β-Ga recovers a more nearly free-electron-like behavior with a DOS at the Fermi energy three times greater than that of α-Ga. A similar recovery of more nearly free-electron-like behavior near the Fermi level occurs in 2D-Ga/SiC.

The Eliashberg spectral function provides more detailed information on the states involved in superconducting pairing in 2D-Ga, from which a theoretical $T_c$ can also be calculated. $\alpha^2F(\omega)$ is converged through a dense Wannier-Fourier interpolated[23] k-point sampling of the electron-phonon matrix elements, the quality of which is verified by comparing the interpolated and directly calculated bands in **Figure 3d** in a ~1 eV window around the Fermi level. **Figure 3e** compares $\alpha^2F(\omega)$ to the projected phonon density of states for the three individual Ga atomic layers and the interfacial Si atom. As shown by the cumulative $\lambda(\omega)$, the dominant contributions to the final $\lambda$=1.62 come from Ga vibrations below 120 cm$^{-1}$, with only modest evidence (near 180 cm$^{-1}$) of qualitatively distinct coupling channels specific to the interfacial Ga and little sign of involvement from the interfacial Si. **Figure 3f** shows the momentum-resolved electron-phonon coupling[24] $\lambda_k$ across the Brillouin zone for electronic states within ±0.5 eV of the Fermi energy. The dominant contributions to $\lambda_k$ come from electron pockets near the two symmetry-inequivalent K points, similar to MoS$_2$ that is n-doped sufficiently to exhibit superconductivity[25]. Further development of CHet-enabled 2D metals with lower lattice symmetry (e.g. ordered alloys) and stronger spin-orbit coupling (e.g. Sn) may be able to access Ising pairing[25] or other exotic states. Using the McMillian-Allen-Dynes formula[26,27] with $\lambda$=1.62 and $\mu^*$ from 0.1 to 0.15 yields a $T_c$ of 3.5 to 4.1 K, in reasonable agreement with experiment. Literature reports of the electron-phonon coupling strength in β-Ga are sparse (and thus somewhat uncertain), but suggest that it is also a reasonably strong-coupled superconductor[28], unlike more weakly coupled α-Ga[26].

The large $\lambda$=1.62 derived from the Ga states alone suggests that the EG layers are unlikely to be the driver of superconductivity in CHet-derived 2D-Ga. Angle-resolved photoelectron spectroscopy (ARPES) of 2D-Ga[15] provides further evidence towards this conclusion: the Fermi level is only 0.2–0.3 eV above the graphene Dirac point, corresponding to n≈8–10×10$^{12}$ cm$^{-2}$, which is 10 to 100× lower than seen in superconducting Li- or Ca-doped epitaxial graphene[29,30]. In those cases, superconductivity is attributed to a partially filled nearly free-electron-like band near the Γ point at a much higher level of charge transfer into graphene than is observed in Ga-intercalated EG. Superconductivity in low-angle twisted bilayer graphene[5] also likely has a very different origin from that in 2D-Ga wherein the required interlayer twist is not present within the EG itself, although the environmental stability of 2D-Ga coupled with the controllable thickness of EG (1-3L post-intercalation) opens prospects to mechanically stack an additional (twisted) graphene monolayer onto EG/2D-Ga to create a hybrid superconducting system. Furthermore, there are obvious prospects to grow thin films of diverse 2D quantum materials directly on the unreactive upper graphene surface of 2D-Ga to create more exotic superconducting heterostructures with high-quality interfaces. Ultimately, the realization of an air-stable 2D single-crystal elemental superconductor at the interface of graphene and SiC further opens the door to stabilizing additional 2D allotropes of conventional 3D superconductors and their alloys across the periodic table with potentially novel properties, all of which are then candidates for incorporation into advanced multi-component heterostructures.

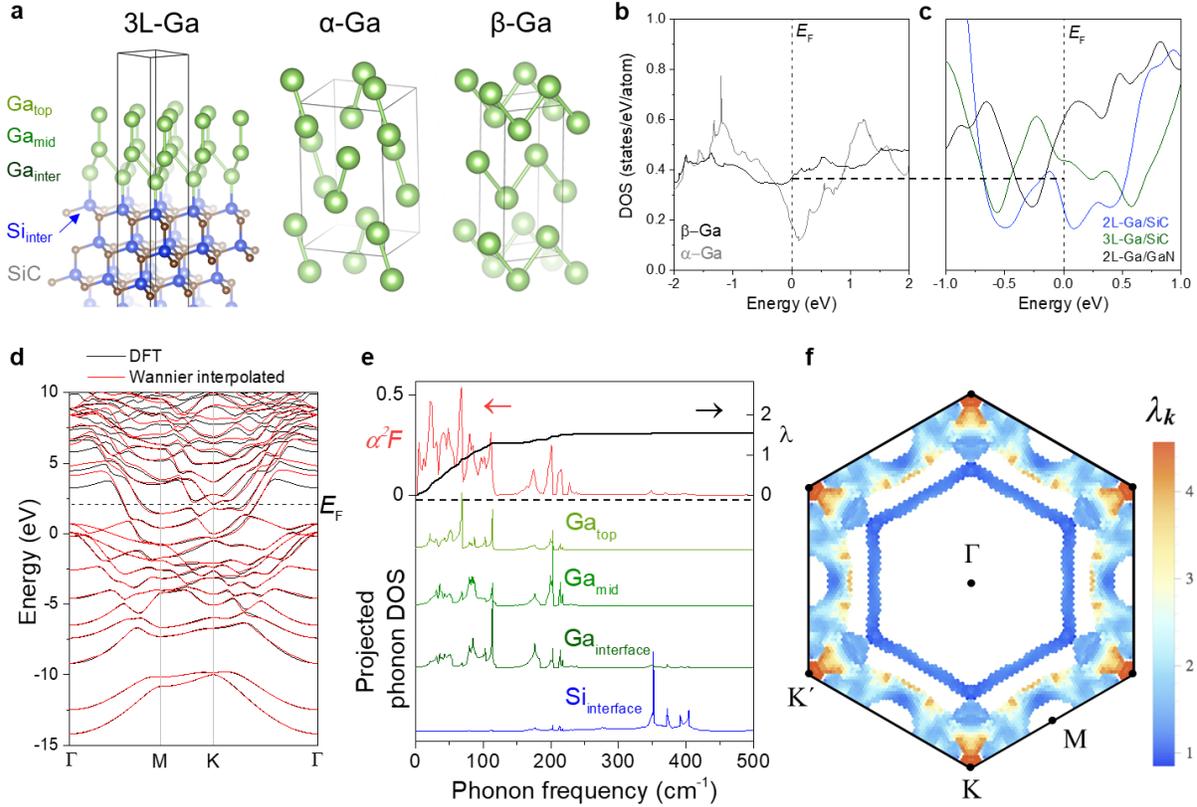

**Figure 3| Theoretical calculations on graphene/2D-Ga heterostructures.** **a**, Atomic structure models of 3L-Ga/SiC, α-Ga, and β-Ga phases. **b**, Electronic density of states (DOS) vs energy density functional theory (DFT) calculations for the two bulk phases of Ga including the stable α-Ga (low-$T_c$) and the metastable β-Ga (high-$T_c$) phases. **c**, DOS vs energy calculations for 2L-Ga/SiC, 3L-Ga/SiC, and 2L-Ga/GaN. Both plots in **b-c** have the same y scales for direct comparison of DOS at the Fermi level and are separated for ease of viewing. **d**, Band structure of 3L-Ga/SiC calculated from DFT (green) and from Wannier interpolation (red) based on the DFT Hamiltonian obtained on a regular 12×12×1 grid. A replication of DFT bands is only required within 0.5 eV of the Fermi level to ensure an accurate estimate of $T_c$. **e**, The Eliashberg spectral function $\alpha^2 F(\omega)$ (orange) compared with the projected phonon DOS of the three types of Ga atoms in trilayer Ga (shades of green) and the top Si atoms at the interface (blue). The top, mid, and interfacial Ga layers are designated in **a**. Importantly, the cumulative electron phonon coupling strength $\lambda(\omega)$ is superimposed in black at the top of **e**. **f**, Momentum-resolved electron-phonon coupling $\lambda_k$ shows that the dominant contribution to coupling strength $\lambda$ comes from the electron pockets near the K and K´ points in the Ga Brillouin zone.

**Methods:**

*Epitaxial graphene growth:*
Epitaxial graphene (EG) is grown via silicon sublimation from the silicon face of 6H silicon carbide (6H-SiC (0001)) in a three-phase, hot-zone, graphite furnace (Thermal Technology LLC). 4H-SiC has also been used successfully for the growth of epitaxial graphene and subsequent intercalation of Gr/Ga heterostructures. The SiC is first annealed in 10% hydrogen (balance argon) at 1500 °C for 30 min to remove subsurface damage due to chemical and mechanical polishing. Then, the $H_2$ is pumped from the system, and the temperature is increased to 1800 °C for 10-30 min at 500-700 Torr to form predominantly 1L graphene plus $(6\sqrt{3} \times 6\sqrt{3})R30°$ interfacial reconstruction layer (buffer layer). Small-step and large-step SiC morphology is largely a factor of SiC wafer miscut but can also be tuned during SiC hydrogen annealing step. The substrate miscut used in this work is specified at ± 0.5° (II-VI, inc.).

*Defect engineering in graphene by plasma treatment:*
Plasma-treatments of epi-graphene substrates is carried out in a PVA Tepla M4L RF gas plasma system using an $O_2$/He (150/50 sccm) gas chemistry at 500 mTorr and 50 W for 1 minute. It should be noted that the M4L is primarily a plasma surface modification system, as opposed to more aggressive reactive ion etchers (or ICP-RIE) tools for deep-etching which tend to completely remove the graphene layers even under the gentlest processing conditions and short etch times. Other plasma chemistries including $CF_4$ gas (via M4L system) and $N_2$:$H_2$ mixtures (via remote plasma in PEALD system) were successful in controllably introducing defects and enhancing intercalation but were not studied in-depth for this work.

*Ga intercalation into epi-graphene:*
Intercalation synthesis is carried out in a STF 1200 horizontal tube-furnace fitted with a one-inch diameter quartz tube. Both the Ga precursor and substrates are loaded into a custom-made alumina crucible from Robocasting Enterprises LLC. Plasma-treated epi-graphene samples are placed face-down (Si-face with 1-2L graphene) several mm above Ga metal precursors (Sigma Aldrich, 99.999%). 40-60 mg of Ga is used per 1x1 $cm^2$ substrate area and are pre-melted by heating the Ga in alumina boat to 50°C for 5 minutes on a hot plate prior to loading into furnace for synthesis. Ga precursor is pre-melted to achieve more uniform evaporation during synthesis. The furnace is heated to 800°C at a ramp rate of 20°C/min, held for 30 min, and then cooled naturally (fan-cooled) to room temperature. The entire process is carried out at 300 Torr under 50 sccm Ar, although higher pressures and lower flow rates have been successfully used as well.

*High-resolution scanning transmission electron microscopy (HR-STEM):*
Cross-section TEM samples are prepared by in-situ lift-out via milling in a FEI Helios NanoLab DualBeam 660 focused ion beam (FIB). Prior to FIB, roughly 60/5/10 nm of $SiO_2$/Ti/Au is deposited via electron-beam evaporation in a Kurt J. Lesker Lab18 evaporator, to improve contrast during STEM imaging at low magnifications. Contrast is improved by increasing the separation distance between graphene/Ga/SiC interface of interest and the bright conductive layers deposited on the sample surface during FIB. Samples are cross-sectioned with Ga+ ion beam at 30 kV then stepped down to 1 kV to avoid ion beam damage to the sample surface. HR-STEM is conducted in a double aberration-corrected FEI Titan[3] G2 60–300 kV S/TEM at 200 kV. Energy dispersive x-ray spectroscopy (EDS) mapping was conducted using the SuperX EDS system under scanning transmission electron microscopy (STEM) mode.

*Auger electron spectroscopy:*
Auger electron spectroscopy mapping is conducted on a Physical Electronics (PHI) Model 670 Scanning Auger system with field-emitter. Maps are acquired at 10 keV using a beam current of 10 nA. Maps are 64x64 pixels corresponding to a pixel size of ~ 0.5 µm. An integration time of 0.1 seconds (per pixel) with 1 eV steps is used. For image quality, 5 frames are averaged for C maps, while 15 frames are averaged for Ga, Si, and O maps. A 2-point acquisition method is used for intensity calculation at each point with the following peak/background energy values used for each of the following elements: (Ga) 1068.0/1080.0 eV, (C) 267.8/291.0 eV, (O) 509.3/532.0 eV, (Si) 1613.2/1628.0 eV.

*Transport measurements and $T_c$, $B_c$, and $T_{BKT}$ extraction:*
Transport measurements were carried out in a Quantum Design physical property measurement system (PPMS) system. Contacts were made to the graphene/Ga heterostructure by lightly scratching the film surface with a diamond scribe or tweezers, and then lightly pressing or soldiering indium dots onto the scratched region. This was done in a casual attempt to make side-contact to the 2D-Ga. Indium dots were arrayed in a standard co-linear four-point-probe configuration with contact pitch on the order of hundreds of microns. All resistance measurements were made with an excitation current of 1 µA. In this work, all critical temperatures are calculated at zero-field, and various $T_c$ parameters are provided in order to facilitate comparisons throughout the literature. $T_c$(onset) is calculated by linearly extrapolating between the transition region and the normal region with the intercept of these two lines defining $T_c$(onset). The linear fit for the transition region is the region of maximum slope, which is pretty consistent for the entire transition width as seen in the log plot inset in **Figure 2a**. $T_c(0.5R_N)$ is defined as the temperature at which the sample reaches half of its normal resistance. $T_c$(zero) is defined as the temperature at which resistance effectively reaches a zero-resistance state i.e. the noise floor of the PPMS system ~ 0.01 Ω. The transition width $\Delta T_c$ is defined as the change in temperature between the $T_c$(onset) and $T_c$(zero). Various $T_c$ values are provided in order to help comparison with other works in literature which may use different values. Critical field $B_{c2}(0.9R_N)$ is defined as the magnetic field at which the sample reaches 90% of its normal resistance. Coherence length is estimated from $B_{c2}(T) = \frac{\Phi}{2\pi\xi_0^2}$. The power-law fitting for the $T_{BKT}$ extraction was done in the lower current and voltage regime (near the bottom of each curve) as we were not able to measure the full I-V curves up to the critical current $I_c$ and into the normal state (following $V \propto I$) due to PPMS current/voltage limitations. Because of this, $T_{BKT}$ is likely higher than 3.14 K if the power-law exponent curve fitting is done closer to $I_c$ where the slope is usually steepest, as is reported in other works.

*Scanning tunneling spectroscopy/microscopy (STS/STM):*
Ga-intercalated graphene/SiC was studied using ultra-high vacuum low-temperature scanning tunneling microscope with *in-situ* out of plane magnetic field at the Center for Nanophase Materials Sciences at Oak Ridge National Laboratory. The sample was preheated to 200 C to remove surface adsorbates at UHV with a base pressure of 2×10-10 Torr before transferring in-situ to STM stage. STM/S was conducted using mechanically cut Pt-Ir tip. All Pt-Ir tips were conditioned and checked using clean Au (111) surface before each measurement. Topographic images were acquired in constant current mode with the bias voltage applied to the samples. All the spectroscopies were obtained using the lock-in technique with bias modulation at 973 Hz. The STM image in Figure 1e was taken at $V_b$ = 10 mV and $I_t$ = 400 pA. The STM image in Figure 1f was taken at $V_b$ = -100 mV, $I_t$ = 100 pA. The dI/dV spectra in Figures 1g-i were measured at $V_b$ = 5 mV, $I_t$ = 400 pA, and $\Delta V$ = 0.1 mV.

*DFT modeling and $T_c$ calculations:*
The electronic density of states calculations are performed using the Vienna Ab-initio Simulation Package (VASP) with the Perdew-Burke-Ernzerhof parametrization of the generalized gradient approximation (GGA-PBE) exchange-correlation functional and projector augmented wave (PAW) pseudopotentials. Seven units of SiC (passivated by hydrogen from below) are included in the structures of 2L- and 3L-Ga/SiC. In preparing Ga/SiC structures, all relaxations are performed using a plane-wave energy cutoff of 400 eV, a k-point grid of 20×20×1, and a force convergence threshold of 0.01 eV/Å. In Figure 3c, bilayer and trilayer Ga on SiC exhibits a DOS at $E_F$ similar to β-Ga (**Figure 3i**), where for bilayer Ga we artificially shift $E_F$ by 0.5 eV to account for the additional (undetermined) electron doping so the band alignment agrees with ARPES measurements[15]. As for the DOS calculations carried out on hexagonal 2L and 3L-Ga/SiC, the "sc" and "scc" stacking sequences were used, respectively, in which 's' stands for Si sites and 'c' stands for C sites, denoting the vertical alignment of the Ga atoms in each layer with respect to the topmost atomic sites in the SiC surface. In this case, the first Ga atom (interfacial Ga) is aligned over the Si atom, while the second and third Ga atoms (Ga middle and Ga top) are aligned over the C atom in SiC. The 'scc' stacking sequence occupies one of the lower energy configurations out of all the possible stacking sequences for 3L-

Ga and most closely matches the band structure as directly measured in ARPES[15]. Thus, 'sc' and 'scc' stackings were used to calculate DOS and the following parameters.

All calculations related to electron-phonon interactions are performed in a cell with only two SiC units, due to the heavy computational demand of these routines; SiC slabs are passivated from below by H atoms with the same mass as Si. The starting-point electronic charge density is calculated on a 12×12×1 Γ-centered k-point grid. Electronic wavefunctions are then computed for a 6×6×1 grid. The phonon dispersion is calculated using density functional perturbation theory based on the same 6×6×1 grid. All computations above are performed by the Quantum ESPRESSO package using the local density approximation exchange-correlation functional, Hartwigsen-Goedeker-Hutter norm-conserving pseudopotentials, and a plane wave expansion cutoff of 1090 eV. To achieve a dense sampling of electron-phonon coupling matrix elements across the Fermi surface, we construct electronic and phonon Wannier functions based on wavefunctions and phonon modes sampled on the coarse 6×6×1 grid and generate interpolations onto a 48×48×1 grid, as implemented by the EPW code. Wannier functions are initialized by projecting the following orbitals onto Bloch wavefunctions: two $s$ and one $p_z$ for each Ga, one $sp^3$ orbital for each Si, and one $sp^3$ for each C. An outer disentanglement window (i.e. one that captures all targeted bands with the chosen orbital characters) coincides with the entire energy range in Fig.3d. An inner window (where all Bloch states are included within the projection manifold) spans the energy range from the lower bound of Fig.3d up to 1 eV above the Fermi level.

The Eliashberg spectral function in the isotropic formalism is given by $\alpha^2 F(\omega) = (1/2N_F) \Sigma_{kqv} |g_{mn}^v(k,k+q)|^2 \delta(\varepsilon_{n,k}) \delta(\varepsilon_{m,k+q}) \delta(\omega-\omega_{qv})$, where $N_F$ is the density of states at the Fermi level, $g_{mn}^v(k,k+q)$ is the el-ph matrix elements characterizing electrons scattering from state $(n,k)$ to state $(m,k+\mathbf{q})$ by a phonon of mode $v$, with their respective energies given by $\varepsilon_{n,k}$, $\varepsilon_{m,k+q}$ and $\omega_{qv}$. The cumulative electron-phonon coupling strength is given by $\lambda(\omega) = 2\int^{\omega} d\omega' \alpha^2 F(\omega')/\omega'$. The variation of electron phonon coupling contributions across the Fermi surface is shown by plotting the momentum-resolved el-ph coupling strength $\lambda_k = \Sigma_{k',v} \delta(\varepsilon_{k'}) |g^v(k,k+q)|^2/\omega_{k-k',v}$ in the Brillouin zone in **Figure 4f**. Lastly, $T_c$ is given by the Mcmillan-Allen-Dynes formula $T_c = \omega_{\log} \exp[-\frac{1.04(1+\lambda)}{\lambda - \mu^*(1+0.62\lambda)}]$, where the logarithmic-averaged phonon frequency $\omega_{\log} = \exp[\frac{2}{\lambda} \int d\omega \log(\omega) \frac{\alpha^2 F(\omega)}{\omega}]$ and $\mu^*$ is the coulomb pseudopotential. See **SI** for more details and for references regarding theoretical el-ph interaction calculations.

**Acknowledgements:**
B.B., N.B., and J.A.R. acknowledge Northrop Grumman Missions Systems university-funded research program, and the Semiconductor Research Corporation Intel/Global Research Collaboration Fellowship. S.S. and J.A.R. acknowledge support from NSF CAREER Award 1453924. J.A.R., V.C., Y.W., N.B. acknowledge the 2D Crystal Consortium National Science Foundation (NSF) Materials Innovation Platform under cooperative agreement DMR-1539916. C.D. acknowledges the Chinese Scholarship Council. C.Z.C. acknowledges support from NSF-CAREER award (DMR-1847811) and an Alfred P. Sloan Research Fellowship. J. Z. is supported by NSF-DMR-1708972. STM/STS study was conducted at the Center for Nanophase Materials Sciences (Oak Ridge National Lab), which is a DOE Office of Science User Facility.

B.B., J.A.R., and V.C. wrote the paper with input from other co-authors. B.B., J.A.R., and Y.W. devised the experiments and modeling approach with input from V.C., and J.Z. B.B. prepared samples, performed materials characterization, and analyzed data. N.B. performed the intercalation growths and assisted with materials characterization. Y.W. performed the modeling under the direction of V.C. J.J. performed electrical measurements under the direction of C.Z.C. K.W. performed cross-sectional STEM imaging.

C.D. and S.S. performed the graphene synthesis under the direction of J.A.R. M.F., Q.Z., G.Z., and A.P.L. performed STM and STS characterization. Y.W.C assisted with electrical measurements under the direction of J.Z. The authors would also like to acknowledge Haiying Wang for help with STEM sample cross-section preparation via FIB, Vince Bojan for help with auger electron spectroscopy, and Max Wetherington for constant Raman spectroscopy system support.

**Supplementary information**

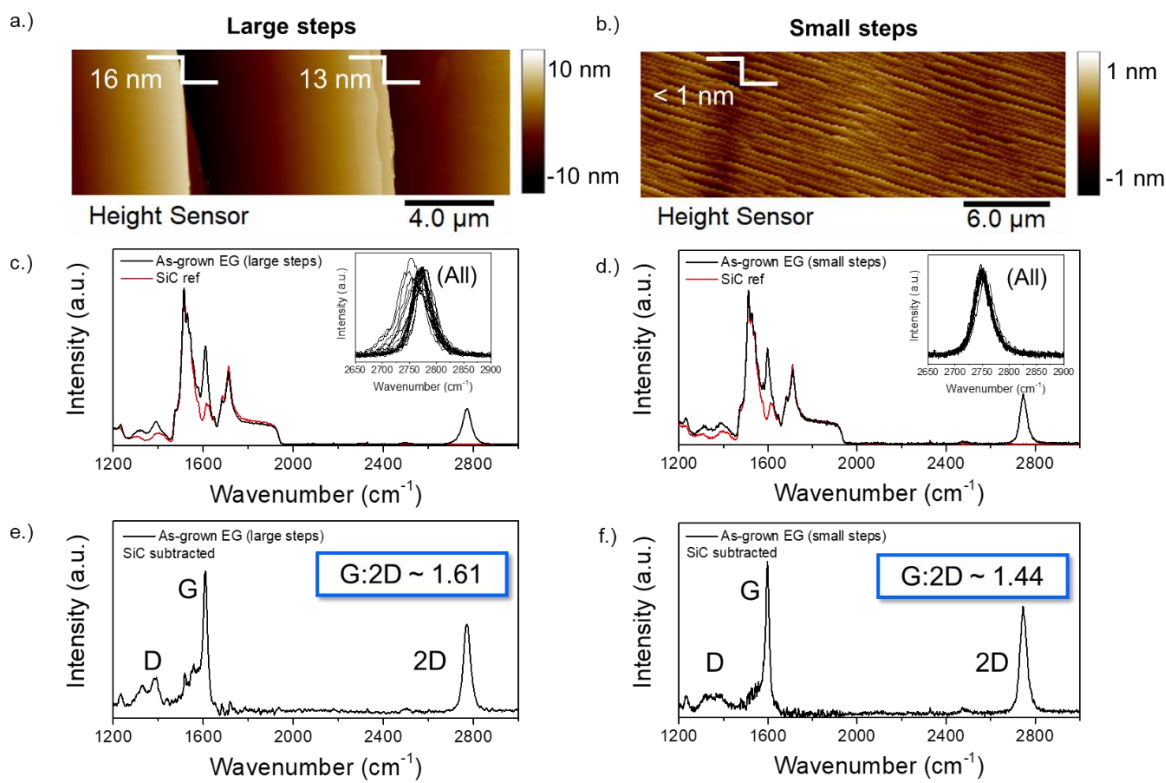

Figure S2: (a-b) Atomic force microscopy (AFM) height images of as-grown epitaxial-graphene (EG)/SiC(0001) for both large step-edge (with step bunching) and small step-edge (no step bunching) morphologies, respectively. (c-d) Raman spectra of EG/SiC and a bare SiC substrate reference sample for both large and small step-edge morphologies. Spectra are normalized to the largest SiC peak ~ 1550 cm$^{-1}$. Insets: all spectra (21 curves) from a 20 µm line scan overlaid, with focus on the 2D peak line shape. The variation in 2D peak line shape and peak position in the large step sample is due to the thicker graphene at step-edges as well heterogeneity in the as-grown graphene within large terraces (strain, etc.). (e-f) Representative individual spectra from as-grown EG with SiC signal subtracted for both step-edge morphologies. In the case of the large step-height sample, the spectra are taken from a flat terrace region as opposed to a step-edge. All graphene used in this work is nominally 1L thick (2L post-intercalation) as demonstrated by the 2D peak line-shape (single Lorentzian) with full-width-at-half-maximum (FWHM) ~ 30 wavenumbers[1–3] and verified directly by TEM imaging post-intercalation (**Figure S4**). For small-step samples, 532 nm laser was used. For large-step samples, 488 nm laser was used. No major differences in Raman spectra were observed for the two laser lines.

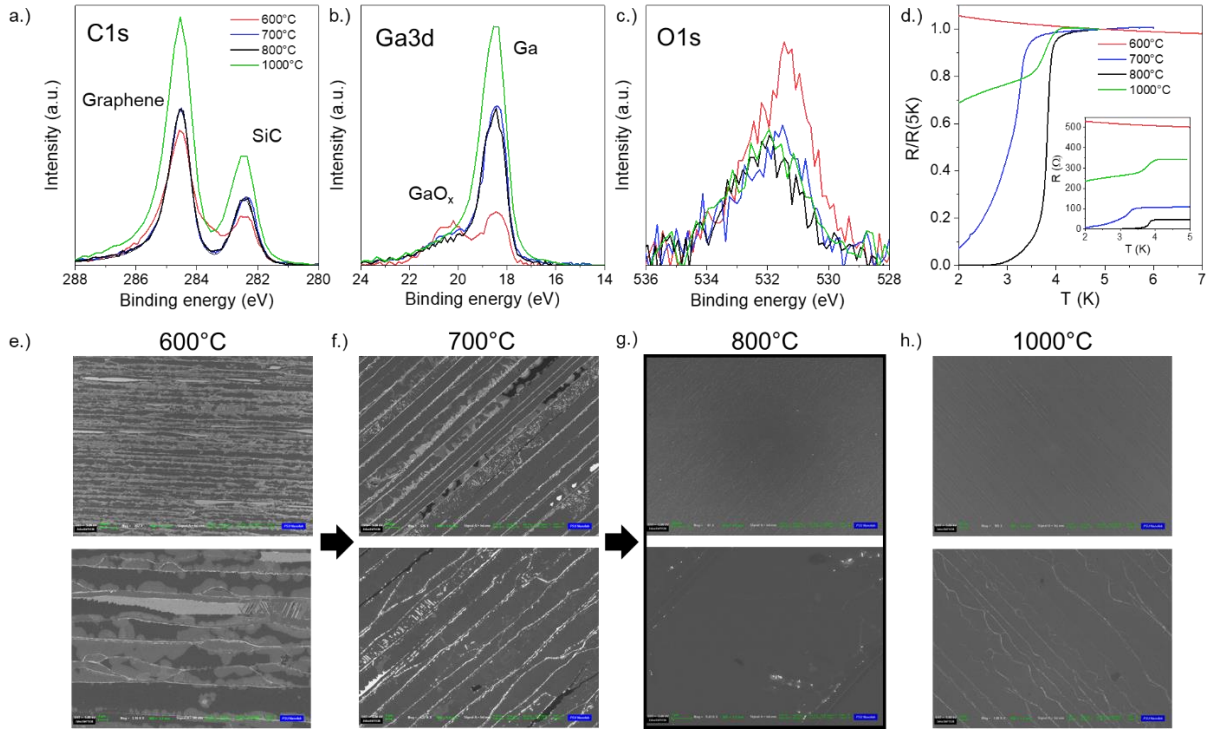

Figure S3: (a-c) C 1s, Ga 3d, and O 1s x-ray photoelectron spectroscopy (XPS) high-resolution spectra of Gr/Ga films intercalated at different temperatures from 600 – 1000°C. For temps ≥ 700°C, there is a large Ga signal and defined C 1s peak splitting[4], suggesting large-area intercalation. All XPS spectra were taken with the same collection conditions and are not normalized to show relative differences in signal (i.e. amount of Ga intercalation) between samples. (d) Zero-field resistance vs temperature (R(T)) curves for Gr/Ga films intercalated at different temperatures from 600 – 1000°C normalized to their normal resistance measured at T = 5 K (above $T_c$). Inset: Non-normalized curves displaying the absolute resistance vs temperature profiles for all intercalation temperatures. (e-h) Scanning electron microscopy (SEM) images of various magnifications for each temperature, showing the relative uniformity in electronic contrast as observed in SEM; there is a clear trend from 600°C to 800°C in increasing uniformity.

The only sample that displays a zero-resistance superconducting transition is the 800°C-film (across many samples). The only film that doesn't display any sign of a transition is the 600°C-film, suggesting a lower temperature cut-off at which Ga intercalation is limited. The other temperatures display varying levels of residual resistance < $T_c$, suggesting varying degrees of intercalated film coverage and connectivity on the macroscale. Interestingly, the 700°C-film displays a lower $T_c$ which suggests some control over Ga-layer thickness and/or phase which may impact superconducting properties, although the lateral extent of intercalation is still limited at 700°C all other conditions held constant. As in the case of 700°C and 1000°C films, XPS alone cannot be used to predict the superconducting properties of the Gr/Ga films. It is unknown what the origins of higher Ga signal are for 1000°C sample. There is low oxygen signal in all films (~2 orders of magnitude lower than Ga3d). Based on these initial measurements, 800°C became the standard temperature used for intercalation studies.

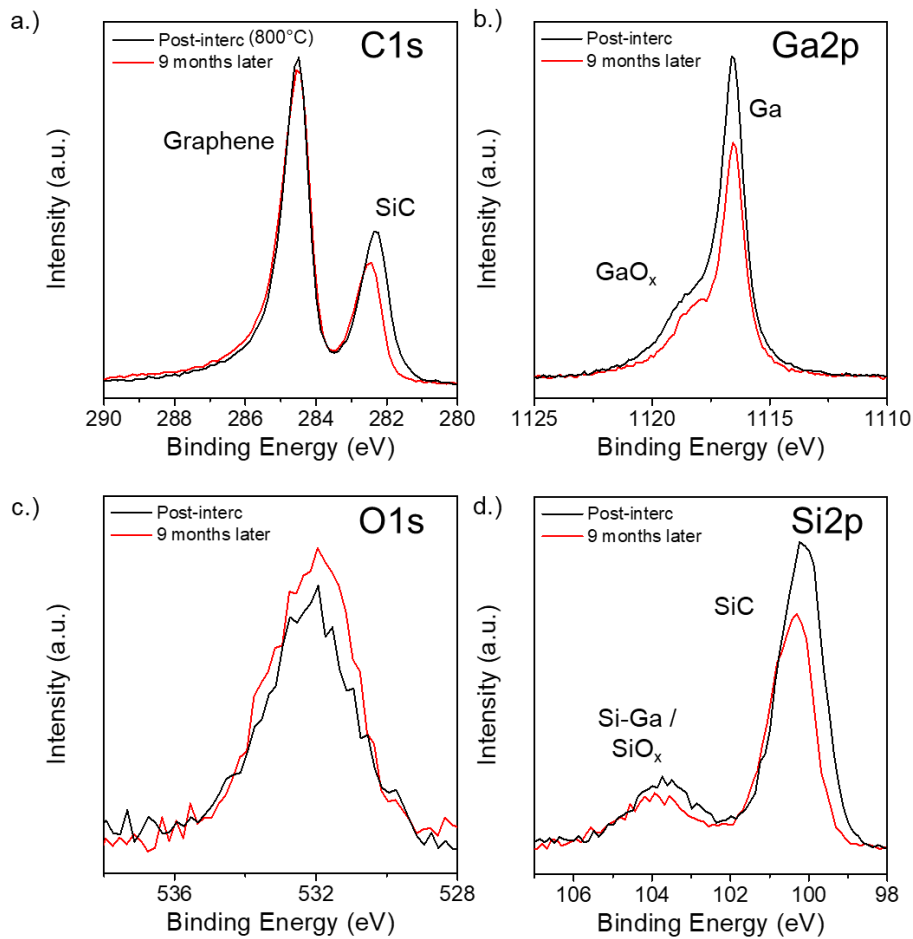

Figure S4: (a-d) x-ray photoelectron spectroscopy (XPS) characterization of Ga-intercalated samples immediately post-growth (within several days) and 9 months after initial growth including C 1s, Ga 2p, O 1s, and Si 2p high resolution spectra, respectively, showing little change or degradation in sample quality, highlighted by little change in bonding peak position, line shape, and overall intensities. Measurements were conducted on different samples so there may be slight differences in spectra and overall intensities due to substrate/sample variation as well as spot-size position. The sample used for "9 months later" sample is one of the samples used for STM/STS characterization (**Figure 1, S8-9**) and routinely was left out in air for days at a time. Most importantly, the oxygen (O 1s) intensity is relatively unchanged after 9 months.

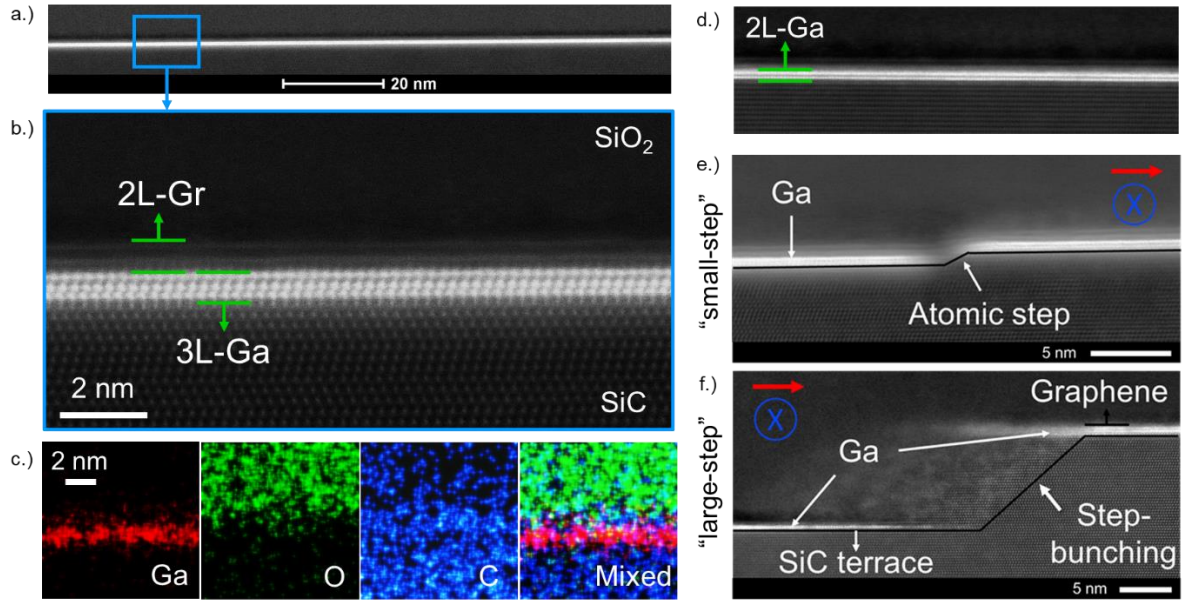

Figure S5: (a) Low-magnification high-resolution scanning transmission electron microscopy (STEM) image of a Gr/Ga/SiC heterostructure showing uniform film height and continuity across > 100 nm of cross-section. (b) The same high-magnification HR-STEM image of a 2L-graphene/3L-Ga heterostructure from **Figure 1a** in the main text. (c) Energy dispersive x-ray spectroscopy (EDS) maps of Ga, O, and C as well as a mixed map with all elements overlaid. Importantly, there is little-to-no oxidation of the 2D-Ga film detected in EDS. (d) A high-magnification HR-STEM image of a 2L-Ga film. 2-3L of Ga is the predominant thickness of Ga observed in 800°C films. Other layer thicknesses ranging from 1-5L have also been observed but are usually only several to tens of nanometers in length and are usually found at step-edges or other disruptions in the SiC. (e-f) More STEM images of the Ga films at step-edges/step-bunching of various heights including a "small-step" sample (e) and a "large-step" sample (f) showing the variation in Ga film separation that can occur across adjacent terraces. In the case of small-step heights (~ 1 nm tall), Ga terraces are in close proximity allowing a superconducting current to tunnel easily. In the "large-step" samples, step-bunching heights can range from 1 – 20 nm in height which introduces finite resistance into our R-T measurements. The red arrow and blue X correspond to the perpendicular and parallel current directions, respectively, that are discussed in **Figure 1j-k** in the main text.

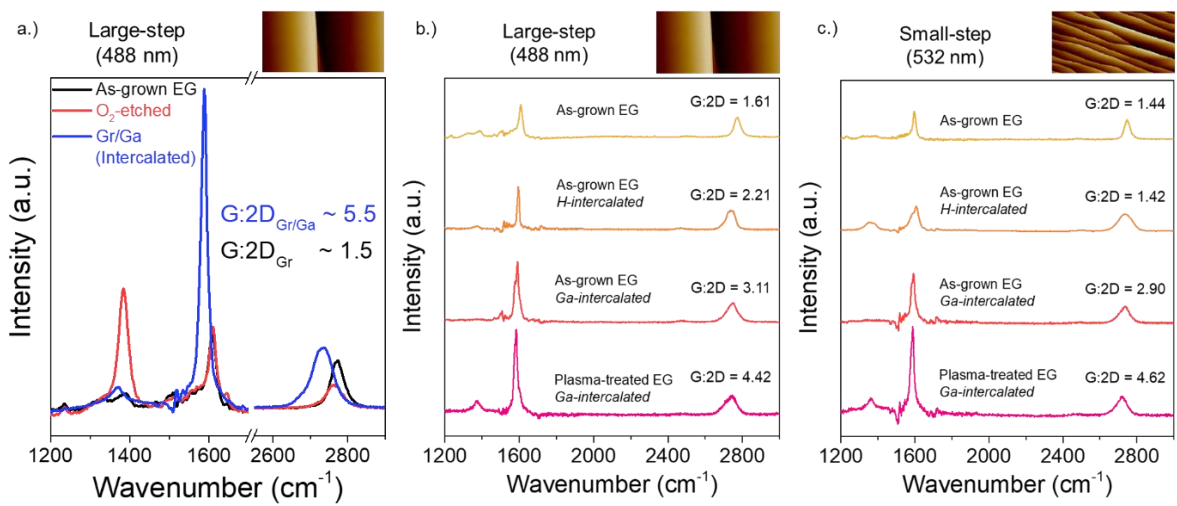

Figure S6: (a) Representative Raman spectra from the 3 major steps in synthesis of Gr/Ga heterostructures showing the evolution in graphene properties including: post graphene-growth, post-$O_2$-plasma treatment, and post intercalation (annealing in Ga). In (a), spectra are normalized to the largest SiC peak ~ 1550 cm$^{-1}$ and then SiC-background subtracted. The two foremost features of this graphene spectral evolution are the almost complete removal of graphene D peak (~1350 cm$^{-1}$) and significant increase in G peak (~1600 cm$^{-1}$) intensity post-intercalation. The graphene used in (a) is large-step graphene/SiC substrates, measured with 488 nm, and curves are overlaid to demonstrate relative changes in spectra between processing steps. (b-c) Representative Raman spectra for both graphene/SiC morphology types studies in this work including large-step (488 nm laser) and small-step (532 nm laser) samples, respectively, for various processing conditions. The spectra shown here include as-grown graphene (non-intercalated), as-grown graphene (H-intercalated), as-grown graphene (Ga-intercalated), and plasma-treated graphene (Ga intercalated). It should be noted that in the case of Ga-intercalated as-grown graphene that has not been plasma-treated, the majority of the surface is not intercalated; the spectra shown in these plots are taken from an intercalated region where Ga is present. Also, H-intercalated graphene does not display as large a G:2D intensity ratio as Ga-intercalated graphene, which suggests this G peak enhancement is due to plasmonic effects from the 2D-metal. Lastly, the plasmonic enhancement of the G peak is not as strong for partial Ga-intercalated islands in as-grown graphene as it is for fully-intercalated samples (where the graphene was plasma-treated). We attribute this to partial strain release, as the graphene is still 'anchored' to the SiC immediately outside of intercalated islands. Residual defects in the plasma-treated graphene layers could also play a role in the plasmonic coupling.

**Raman discussion:**

**Figure S5a** displays representative Raman spectra after the three major steps of the EG/Ga synthesis process: post EG growth (black), post $O_2$-plasma treatment (red), and post Ga intercalation (blue). Analysis of the D (~1350 cm$^{-1}$), G (~1600 cm$^{-1}$), and 2D (~2750 cm$^{-1}$) graphene peaks provide evidence of the relative defectiveness (D intensity, D:G intensity ratio), layer thickness (2D peak width), doping/plasmonic enhancement (G:2D intensity ratio), and strain (G, 2D peak shifts)[1–3,5,6], making Raman spectroscopy an invaluable tool for graphene characterization. The as-grown EG used in this study is nominally 1L thick (plus buffer) pre-intercalation and 2L thick post-intercalation due to buffer layer-to-graphene conversion as confirmed in TEM and the evolution in Raman spectroscopy. The G:2D intensity ratio for as-grown EG is on average ~1.5, with a 2D peak full-width-at-half-maximum (FWHM) ~30 cm$^{-1}$, indicating monolayer[3]. Based on the D:G ratio, the Raman spectra post-plasma treatment confirms a significant increase in defect density (~200×; D:G > 1)) and increase in C-O and C=O bonding detected via XPS[7]. The 2D peak intensity is also decreased post-plasma-treatment due to interruption in the sp$^2$ bonding in graphene.

Ga-intercalation greatly modifies the graphene layer properties, as measured by the Raman spectral evolution between all processing steps. First, there is a broadening and slight intensity increase of the 2D peak, indicative of an increase in the graphene thickness by 1L due to the buffer layer conversion. There is also a downshift in both the G and 2D peak positions, which indicates a release of the built-in compressive strain inherent in as-grown EG on SiC[1]. In addition to buffer-layer conversion, there is a significant reduction in the defect density post-intercalation, which is evidenced by the overall decrease in the D peak intensity back to as-grown levels and accompanying decrease in the D:G ratio (from ≥ 1 post-plasma to < 0.1 post-intercalation). While the defects are essential for enhanced, uniform intercalation, as discussed in the section, they are undesirable afterwards if the graphene is to serve as an effective encapsulation layer. This observed "defect healing" is most likely a result of the 800°C anneal as well as the reported catalytic properties of Ga for graphene growth[8]. Lastly, the most striking feature is a significant (3-4×) increase in the G peak intensity for Ga-intercalated regions leading to a G:2D peak intensity ratio (G:2D) ~ 5.5, much higher than H-intercalated (G:2D ~ 2.5) graphene (**Figure S5 b-c**). Furthermore, while doping density changes can impact the G-band intensity,[5,6] recent ARPES data suggests that the doping levels of as-grown EG and Ga-intercalated EG are similar (8 - 10 x$10^{12}$ cm$^{-2}$)[7]. Thus, we attribute the G peak intensity enhancement to surface plasmon enhancement, similar to that observed for Ga particle decorated graphene[9,10] and plasmonic graphene/metal nanostructures[11]. Ultimately, the G-band enhancement with Ga-intercalation enables one to utilize the G:2D ratio for spatially identifying 2D-Ga in addition to complementary optical and electronic image contrasts.

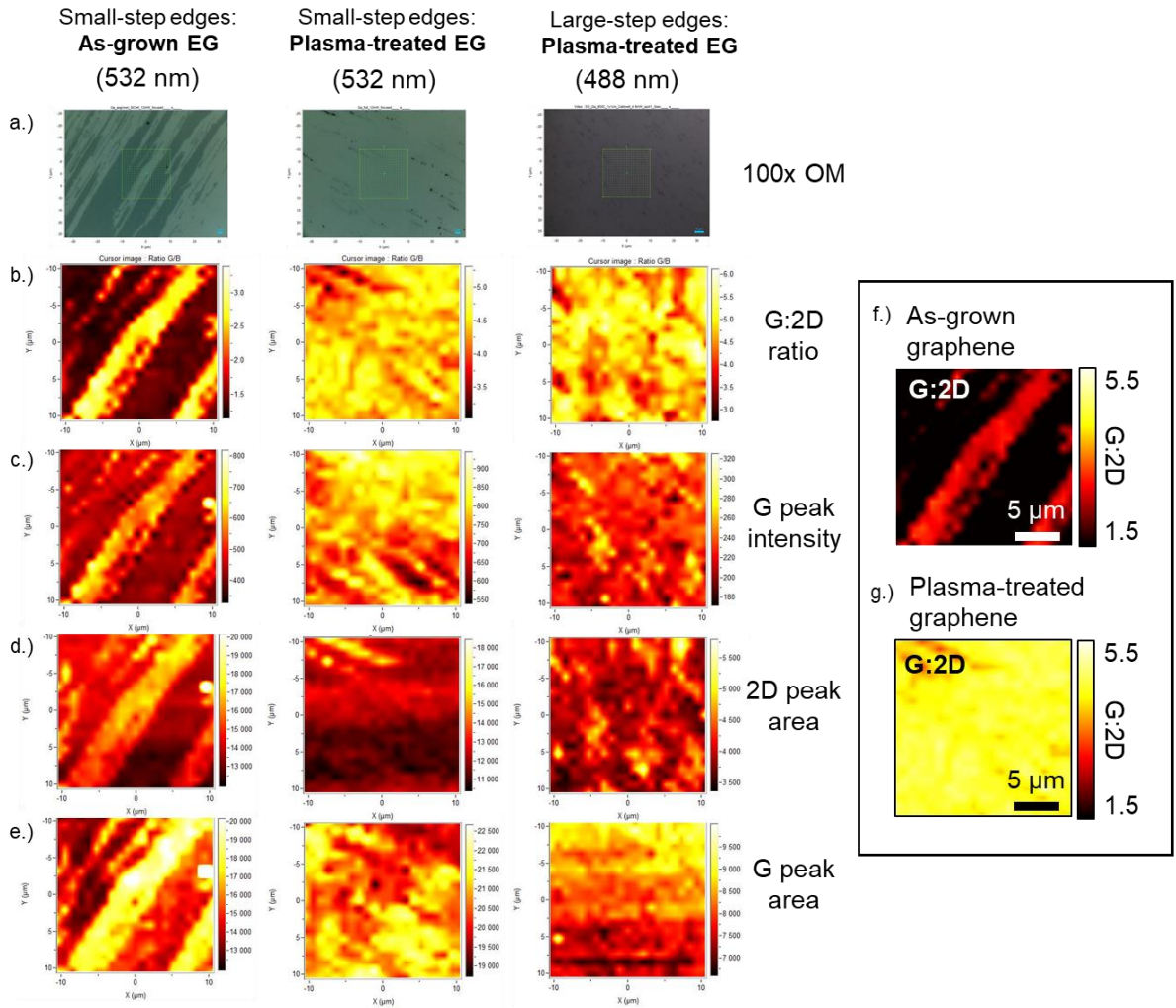

Figure S7: (a) Optical microscope (OM) images of 3 intercalated Gr/Ga films synthesized from different graphene types including as-grown graphene (small step morphology), plasma-treated graphene (small step morphology), and plasma-treated graphene (large step morphology). The green box in the center of the OM images indicates where the following 20x20 µm Raman maps were collected: (b) G:2D peak intensity ratio, (c) G peak intensity, (d) 2D peak area-under-the-curve, and (e) G peak area-under-the-curve maps. (f-g) G:2D ratio maps for as-grown and plasma-treated small-step samples normalized and given the same color scale to illustrate the differences in G:2D ratio in these two cases. All other maps in this figure are automatically normalized individually in the Horiba Labspec software based on their individual intensity count histograms. All scale bars and intensities are shown. The 3 separate mappings shown here are all 20x20 µm with 1 µm step-size (pixel size). 532 nm laser (~12 mW) was used for the first two maps of small-step epitaxial graphene. 488 nm (~5 mW) was used for the last map of large-step epitaxial graphene. All graphene used in this work regardless of SiC step-edge morphology is nominally 1L thick. The Raman maps can be summarized in the following few sentences. For all graphene types, the G:2D ratio is the best means for identifying intercalated regions (in addition to optical and electronic contrasts). Intercalated regions also display increased full-width-half-maximum (FWHM) / area under the curve for both G and 2D peaks compared to as-grown graphene. It should be noted that significantly large step-edges display increased G and 2D peak intensities and FWHM values due to locally thicker graphene. Small-step-edges are almost imperceptible by Raman mapping.

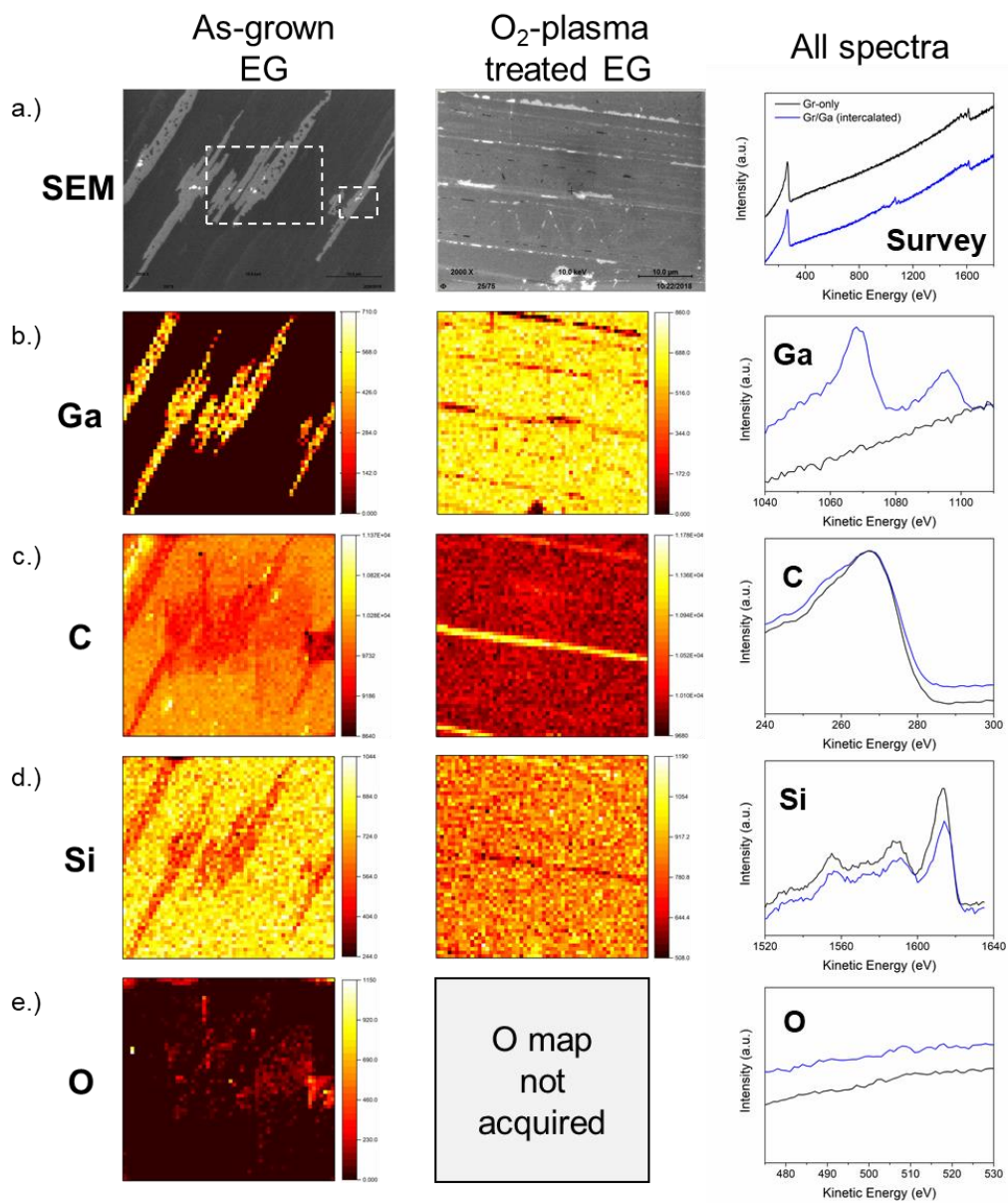

Figure S8: (a) Scanning electron microscope (SEM) images of intercalated Gr/Ga heterostructures synthesized from two different graphene types: as-grown graphene (non-plasma treated) and $O_2$-plasma-treated graphene, respectively. (b-e) Auger electron spectroscopy (AES) elemental maps of Ga, C, Si, and O for both samples. O scan was not taken at the time for the fully intercalated sample via plasma-treated graphene; however, there is little oxygen in either film as evidenced by AES spectra, XPS characterization, and EDS mapping in STEM. The dotted white boxes outlined in the top left SEM image of as-grown graphene are areas that were imaged at higher-magnification in the SEM prior to acquiring lower-magnification AES maps. Imaging at higher mags in the SEM (thus hitting the surface with higher flux of electrons) seems to impact the adventitious carbon make-up on the surface, leading to some pattern transfer in the C and O maps for the as-grown graphene case. The accompanying individual AES spectra for intercalated and non-intercalated regions from as-grown sample (which has both regions present) are given for all elements to the right of the maps.

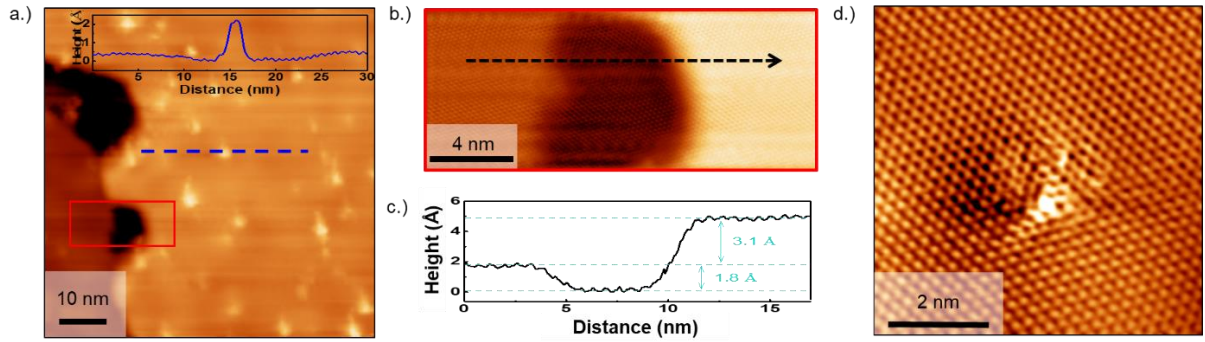

Figure S9: (a) A scanning tunneling microscopy (STM) image of a superconducting Gr/Ga heterostructure film ($V_b$ = 10 mV and $I_t$ = 400 pA). Inset at the top of (a) is a height profile of the dashed blue line intersecting a 2 Å tall defect observed in superconducting terrace regions. The origin of these "island" defects is unknown. (b) A zoomed-in STM image of a step-edge region ($V_b$ = -200 mV, $I_t$ = 200 pA) where STS spectra and superconducting energy gaps were measured. (c) A height profile of the dashed black line in (b), showing relatively small step-height of several angstroms. (d) High-mag STM image of the graphene surface with a carbon vacancy in the center, as evidenced by the triangular distortions around the center ($V_b$ = -100 mV, $I_t$ = 100 pA).

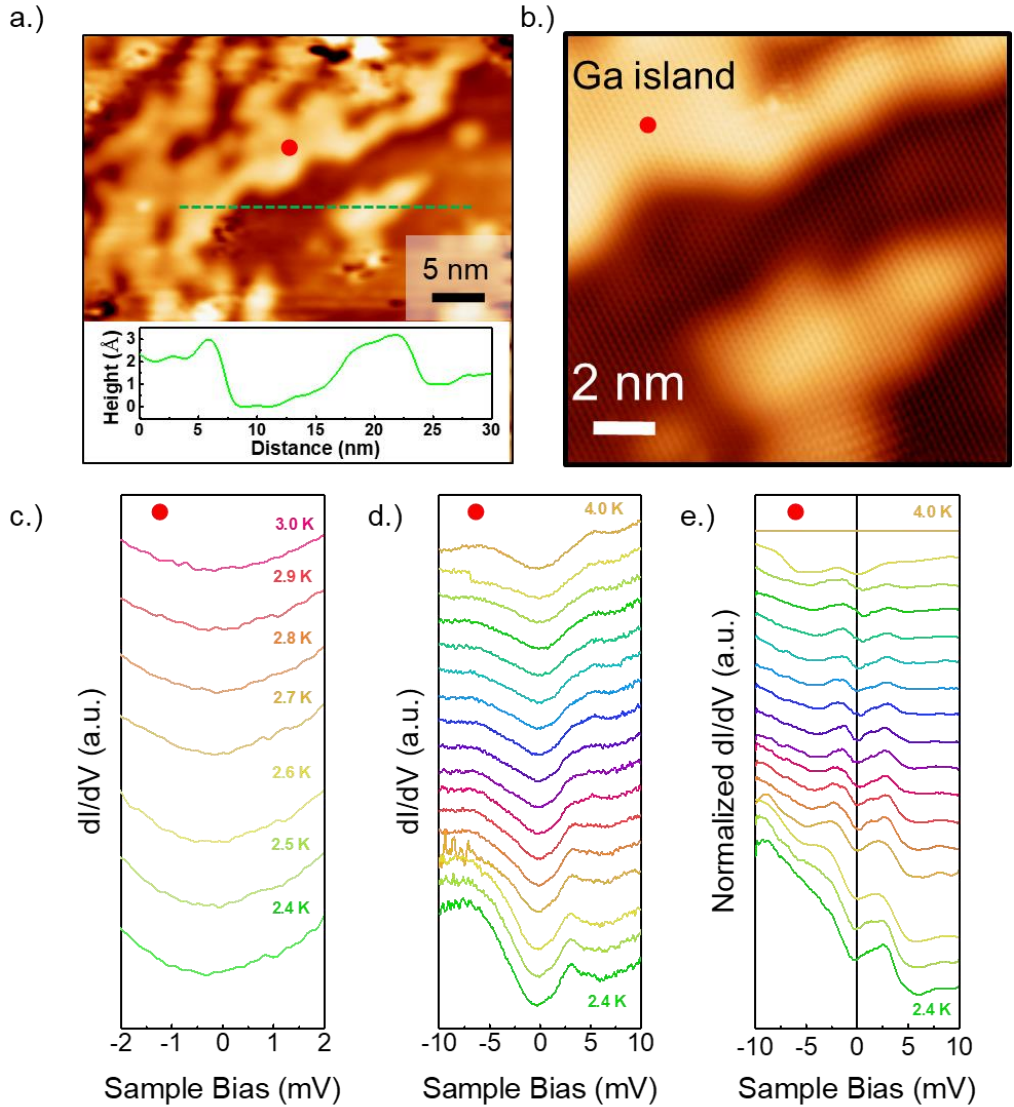

Figure S10: (a) Scanning tunneling microscopy (STM) image of a partially intercalated Gr/Ga film ($V_b$ = 1.0 V, $I_t$ = 400 pA). Inset: height profile corresponding to the dashed green line. (b) Zoomed-in STM image of the area shown in (a) ($V_b$ = 1.0 V, $I_t$ = 100 pA). The graphene lattice is observed across the image, indicating the topography is due to partial Ga intercalation (possible 1 Ga atom) underneath the graphene. (c-d) Temperature dependent dI/dV spectrum at the red mark in (a-b) ($V_b$ = 10 mV, $I_t$ = 400 pA, and $\Delta V$ = 0.1 mV) for different energy ranges. Note that the spectra in (d) are the same as in (c) but over a wider bias range and curves from 3 – 4 K are also displayed for thoroughness. No clear superconducting energy gap with coherence peaks is observed for partially intercalated Ga islands. (e) dI/dV spectra in (d) normalized by 4.0 K curve. We suspect that these Ga islands are non-superconducting due to the greatly reduced domain size compared to the calculated coherence length (~50 nm at 2 K) for 2D-Ga. Such suppression of the superconducting state has been shown for atomically thin Pb islands when their lateral length is reduced well below the coherence length[12].

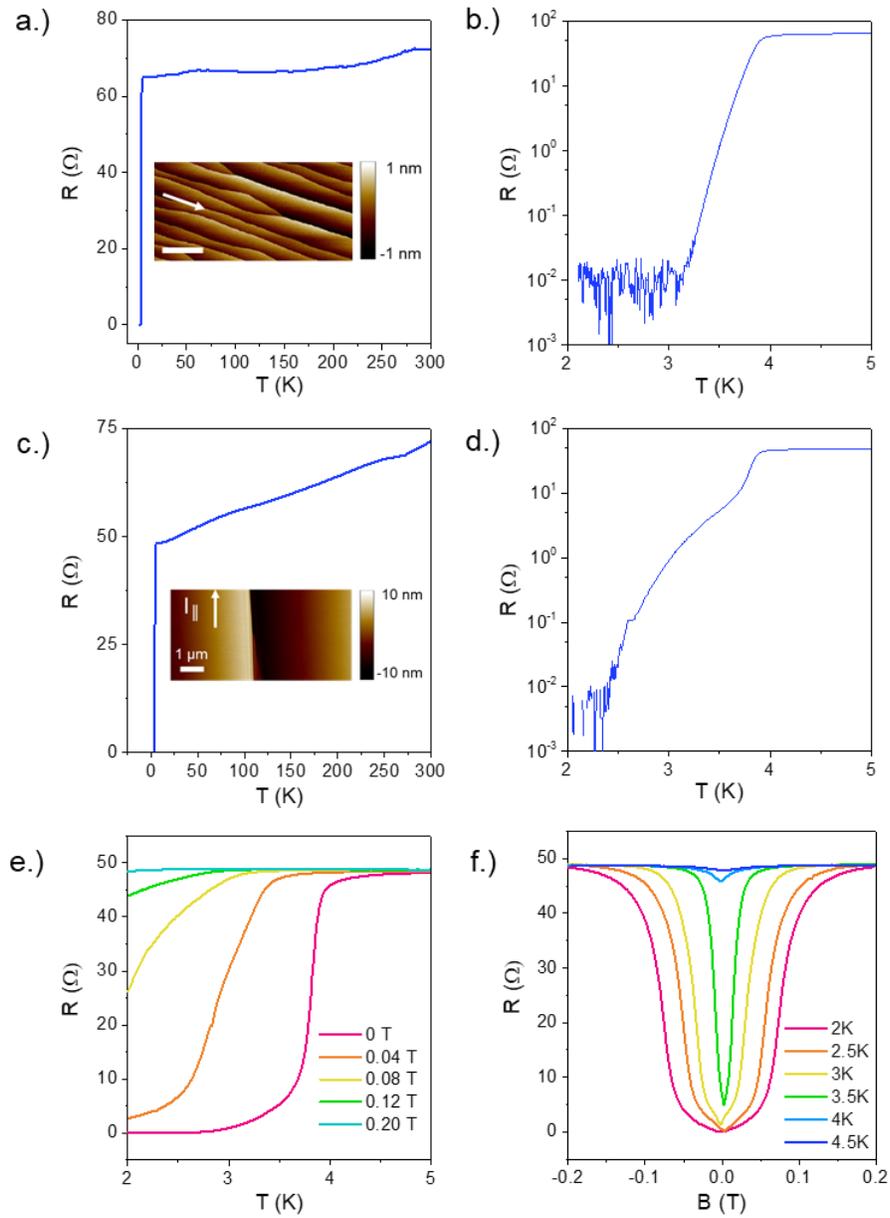

Figure S11: (a,c) Zero-field R(T) curve from 2 – 300 K for a Ga-intercalated sample using small-step and large-step EG/SiC, respectively, displaying largely metallic behavior and a full superconducting transition at low T. Inset: AFM images of EG/SiC morphology in question. (b,d) Log-scale R(T) curves from 2-5 K for small-step and large-step EG/SiC, respectively. These are the same curves in (a,c). (e) B-dependent R(T) curves showing a breakdown in the superconducting phase with increasing perpendicular magnetic field. (f) T-dependent R(B) curves showing a breakdown in the superconducting phase with increasing temperature. (e-f) Superconducting data is taken from large-step sample. Critical temperature and critical fields for large-step sample are similar to that of small-step sample data. All measurements here were taken in the parallel measurement configuration.

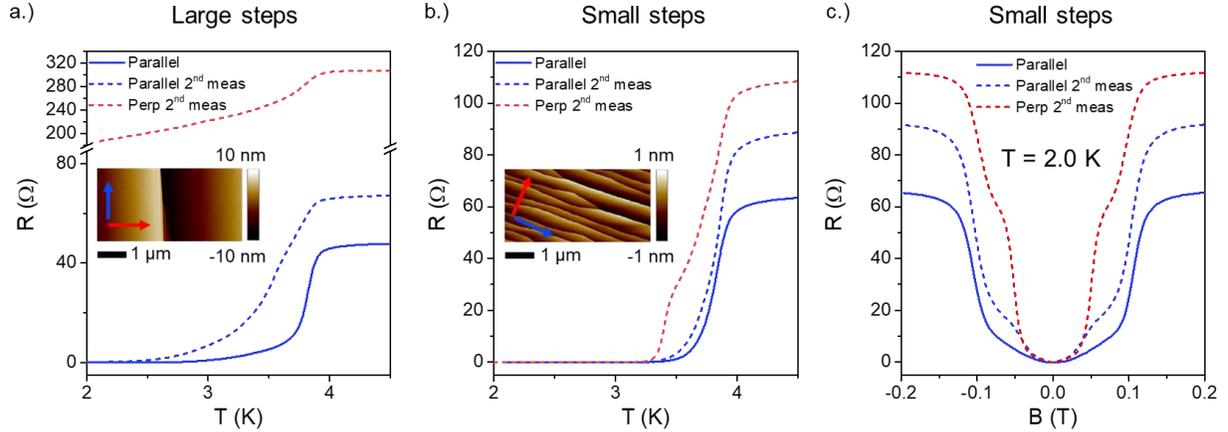

Figure S12: (a-b) Zero-field R(T) measurements for large-step and small-step samples, respectively, showing the difference in transport between subsequent measurements. (c) R(B) curves taken at 2.0 K showing the difference in transport between subsequent measurements, for a small-step sample. For both the R(T) and R(B) curves, resistance is increased for the second measurement (dashed blue line). In addition, the existence of two critical fields becomes more apparent during second set of R(B) measurements, indicating the existence of two distinct superconducting phases. This is likely either 2L vs 3L-Ga regions or terrace vs step-edge regions, in which the local structure of the Ga may differ. In the case of these measurements, samples are completely removed from the PPMS system between measurements. We suspect film degradation is primarily due to oxidation at exposed EG/Ga edges where films were scratched prior to indium dot placement. The dashed curves (second set of measurements) are the same curves in **Figure 1 j-k.** The second set of measurements were used in **Figure 1j-k** in order to make a fair comparison between parallel and perpendicular configurations, as perpendicular measurements were not made on the first sample load and measurement.

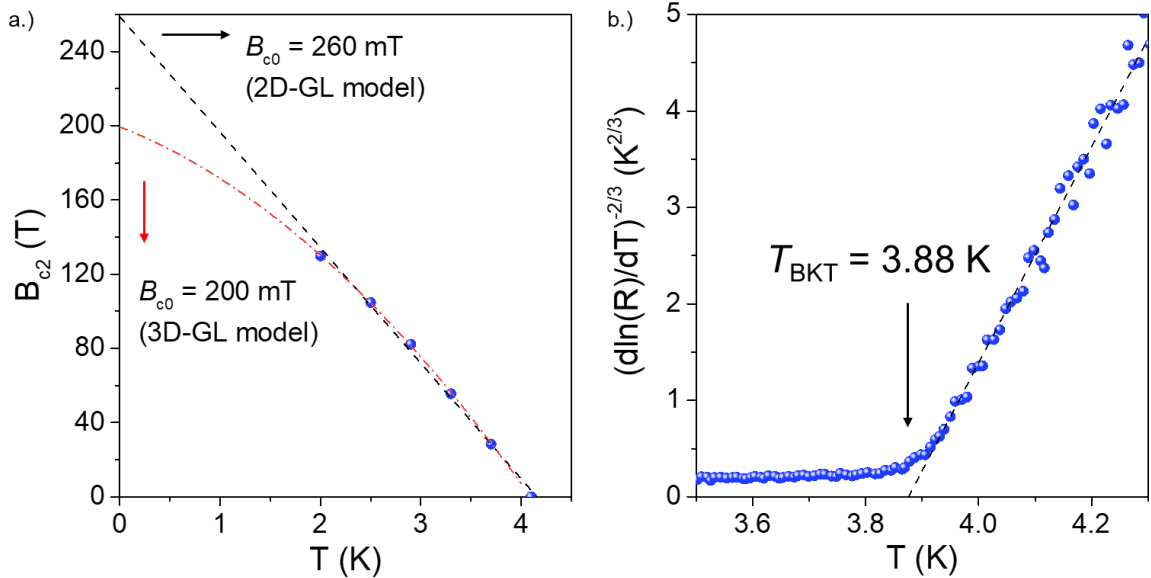

Figure S12: (a) Plot of critical field ($B_{c2}$) vs temperature, extracted from B(T) data **Figure 2c**. Here, $B_{c2}$ is defined as the critical field at 90% of the normal resistance. The data is fitted to the phenomenological 2D Ginzburg–Landau (GL) model: $Bc2(T) = Bc0(1 - \frac{T}{Tc})$, giving a 0 K critical field $B_{c0} \approx 260$ mT. If fitted

with a 3D-GL parabolic relationship $Bc2(T) = Bc0\left(1 - \left(\frac{T}{Tc}\right)^2\right)$, $B_{c0} \approx 200$ mT. (b) Plot of $[d(\ln R)/dT]^{-2/3}$ vs T, showing the extrapolated BKT transition ($T_{BKT}$) temperature from R(T) measurements, specifically the zero-field curve in **Figure 2b**.

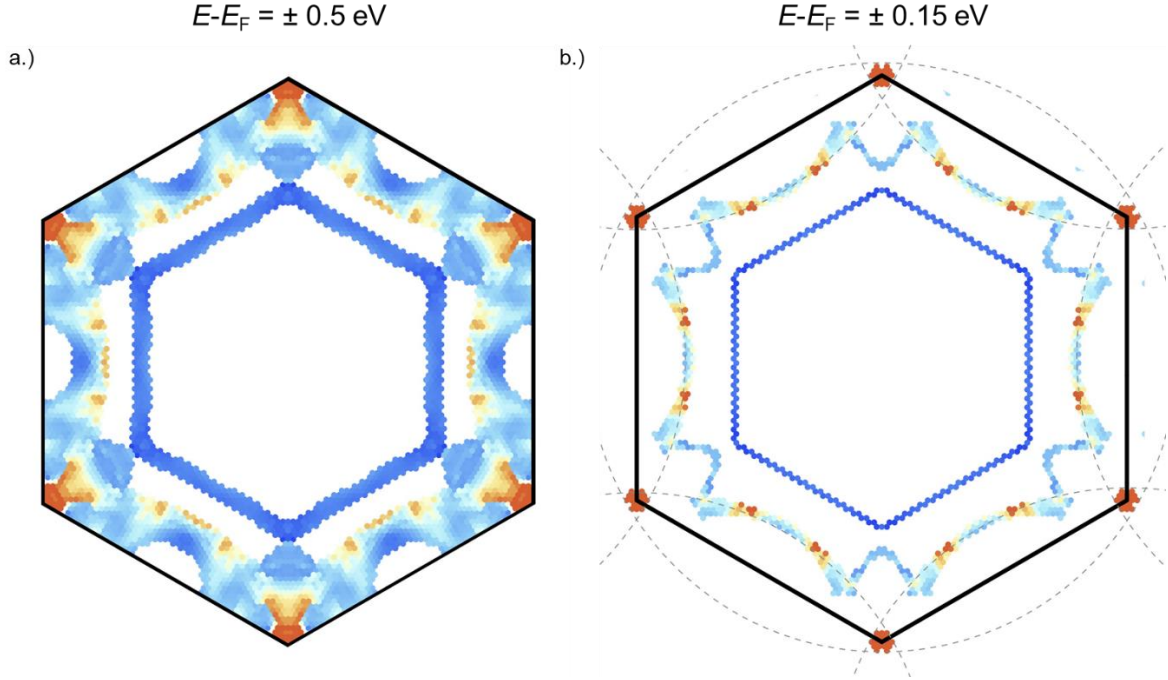

Figure S13: (a-b) Momentum-resolved electron-phonon coupling $\lambda_k$ map for states within ± 500 meV (from **Figure 1e**) and ± 150 meV from the Fermi level, respectively. See **SI Methods** for computational details and procedure.

**Methods:**

*Atomic force microscopy (AFM):*

All AFM micrographs were taken on a Bruker Dimension Icon I/II AFM system at a scan rate of 0.5 Hz and 512 lines per image resolution.

*Raman spectroscopy:*

Raman spectra are acquired on a Horiba LabRam system with 488 nm (532 nm) excitation wavelengths at incident laser power of ~5.0 mW (~ 12 mW). No major differences in spectra were observed between the two laser lines; rather, the Raman line used was based on availability of the respective lines and the individual Raman systems. All Raman data is designated either in figures or captions as to whether 488 or 514 nm was used. A 600 grating/mm filter is used for all measurements. An acquisition time of 10-30 seconds is used for individual point spectra, while 5-10 seconds is used for mapping.

*X-ray photoelectron spectroscopy (XPS):*

XPS measurements were carried out with a Physical Electronics Versa Probe II equipped with a monochromatic Al Kα X-ray source (hv=1486.7 eV). High resolution spectra were obtained over an analysis area of 200 μm at a pass energy of 29.35 eV for C 1s, Si 2p, Ga 3d, and Ga 2p regions, while O 1s regions were collected with a pass energy of 46.95 eV. Spectra were charge referenced to this graphene peak in C 1s corresponding to 284.5 eV.

*Energy dispersive X-ray spectroscopy (EDS):*

HR-STEM is conducted in a double aberration-corrected FEI Titan$^3$ G2 60–300 kV S/TEM at 200 kV. Energy dispersive x-ray spectroscopy (EDS) mapping was conducted using the SuperX EDS system under scanning transmission electron microscopy (STEM) mode.

*DFT modeling and $T_c$ calculations:*

To estimate $T_c$ of 2D-Ga from first-principles, we calculate electron-phonon (el-ph) coupling strength λ derived from the Eliashberg spectral function $α^2F(ω)$ for a 3L-Ga/SiC system without the graphene cap. Achieving a converged Eliashberg spectral function relies on a dense k-point sampling of el-ph matrix elements, as realized using Wannier-Fourier interpolation[13]. The variation of el-ph coupling contributions across all electronic states within ±0.5 eV of the Fermi surface is shown by plotting the momentum-resolved el-ph coupling strength[14] ($λ_k$) in the Brillouin zone (**Figure 3f**). Finally, using the McMillian-Allen-Dynes formula[15,16] with λ=1.62 and $μ^*$ in a range of 0.1 – 0.15 yields a $T_c$ of 3.5 – 4.1 K, in agreement with the experimental measurements, compared to λ = 0.97 and $T_c$ = 5.9 K for β-Ga[17] and λ = 0.40 and $T_c$ = 1.08 for α-Ga[15].

The electronic density of states calculations are performed using the Vienna Ab-initio Simulation Package (VASP)[18] with the Perdew-Burke-Ernzerhof parametrization of the generalized gradient approximation[19,20] (GGA-PBE) exchange-correlation functional and projector augmented wave (PAW) pseudopotentials[21,22]. Seven units of SiC (passivated by hydrogen from below) are included in the structures of 2L- and 3L-Ga/SiC. In preparing Ga/SiC structures, all relaxations are performed using a plane-wave energy cutoff of 400 eV, a k-point grid of 20×20×1, and a force convergence threshold of 0.01 eV/Å. In Figure 3c, bilayer and trilayer Ga on SiC exhibits a DOS at $E_F$ similar to β-Ga (**Figure 3i**), where for bilayer Ga we artificially shift $E_F$ by 0.5 eV to account for the additional (undetermined) electron doping so the band alignment agrees with ARPES measurements[7]. As for the DOS calculations carried out on hexagonal 2L and 3L-Ga/SiC, the "sc" and "scc" stacking sequences were used, respectively, in which 's' stands for Si sites, 'c' stands for C sites, denoting the vertical alignment of the Ga atoms in each layer with respect to the topmost atomic sites in the SiC surface. In this case, the first Ga atom (interfacial Ga) is aligned over the Si atom, while the second and third Ga atoms (Ga middle and Ga top) are aligned over the C atom in SiC. The 'scc' stacking sequence occupies one of the lower energy configurations out of all the possible stacking sequences for 3L-Ga and most closely matches the band structure as directly measured in ARPES[7]. Thus, 'sc' and 'scc' stackings were used to calculate DOS and the following parameters.

All calculations related to electron-phonon interactions are performed in a cell with only two SiC units, due to the heavy computational demand of these routines; SiC slabs are passivated from below by H atoms with the same mass as Si. The starting-point electronic charge density is calculated on a 12×12×1 Γ-centered k-point grid. Electronic wavefunctions are then computed for a 6×6×1 grid. The phonon dispersion is calculated using density functional perturbation theory based on the same 6×6×1 grid. All computations above are performed by the Quantum ESPRESSO package[23] using the local density approximation exchange-correlation functional, Hartwigsen-Goedeker-Hutter norm-conserving pseudopotentials[24], and a plane wave expansion cutoff of 1090 eV. To achieve a dense sampling of electron-phonon coupling matrix elements across the Fermi surface, we construct electronic and phonon Wannier functions based on wavefunctions and phonon modes sampled on the coarse 6×6×1 grid and generate interpolations onto a

48×48×1 grid, as implemented by the EPW code[13,25]. Wannier functions are initialized by projecting the following orbitals onto Bloch wavefunctions: two $s$ and one $p_z$ for each Ga, one $sp^3$ orbital for each Si, and one $sp^3$ for each C. An outer disentanglement window[26] (i.e. one that captures all targeted bands with the chosen orbital characters) coincides with the entire energy range in Fig.3d. An inner window (where all Bloch states are included within the projection manifold[26]) spans the energy range from the lower bound of Fig.3d up to 1 eV above the Fermi level.

The Eliashberg spectral function, in the isotropic formalism, is given by $\alpha^2 F(\omega) = (1/2N_F) \Sigma_{kq\nu} |g_{mn}^\nu(\bm{k},\bm{k+q})|^2 \delta(\varepsilon_{n,\bm{k}}) \delta(\varepsilon_{m,\bm{k+q}}) \delta(\omega-\omega_{q\nu})$, where $N_F$ is the density of states at the Fermi level, $g_{mn}^\nu(\bm{k},\bm{k+q})$ is the el-ph matrix elements characterizing electrons scattering from state $(n,\bm{k})$ to state $(m,\bm{k+q})$ by a phonon of mode $\nu$, with their respective energies given by $\varepsilon_{n,\bm{k}}$, $\varepsilon_{m,\bm{k+q}}$ and $\omega_{q\nu}$. The cumulative electron-phonon coupling strength is given by $\lambda(\omega) = 2\int^\omega d\omega'\, \alpha^2 F(\omega')/\omega'$. The variation of electron phonon coupling contributions across the Fermi surface is shown by plotting the momentum-resolved el-ph coupling strength[14] $\lambda_{\bm{k}} = \Sigma_{\bm{k'},\nu} \delta(\varepsilon_{\bm{k'}}) |g^\nu(\bm{k},\bm{k+q})|^2/\omega_{\bm{k-k'},\nu}$ in the Brillouin zone in **Figure 4f**. Lastly, $T_c$ is given by the Mcmillan-Allen-Dynes formula $T_c = \omega_{\log} \exp[-\frac{1.04(1+\lambda)}{\lambda-\mu^*(1+0.62\lambda)}]$, where the logarithmic-averaged phonon frequency $\omega_{\log} = \exp[\frac{2}{\lambda}\int d\omega \log(\omega) \frac{\alpha^2 F(\omega)}{\omega}]$ and $\mu^*$ is the coulomb pseudopotential.